# Impact of fast ions on turbulent transport in high-β HL-2A scenarios

Jingchun Li[1]*, Zhaoyang Lu[1], Jianqiang Xu[2], Wei Chen[2], Jiaqi Dong[2], Jingting Luo[1], Yong Liu[1]

[1]Shenzhen Key Laboratory of Nuclear and Radiation Safety, Institute for Advanced Study in Nuclear Energy & Safety, College of Physics and Optoelectronic Engineering, Shenzhen University, Shenzhen 510640, People's Republic of China
[2]Southwestern Institute of Physics, PO Box 432, Chengdu 610041, People's Republic of China

**ABSTRACT**. The fast-ion (FI) on turbulent transport is one of the key topics of magnetic confinement fusion. This work focus on the impact of FI pressure gradients on turbulence in a high-β plasma scenario using gyrokinetic simulations. Linear analyses reveal that FIs strongly stabilize ion temperature gradient (ITG) modes via the thermal-ion dilution, while their influence on trapped electron modes (TEMs) is minimal. At elevated FI pressure gradients, a transition to a FI-driven BAE (FI-BAE) regime occurs, as evidenced by mode structure and frequency alignment within the Alfvénic gap. Electron β scans further demonstrate the emergence of kinetic ballooning modes (KBMs) at higher β, whereas an ITG-TEM hybrid turbulence dominates near experimental β values. Nonlinear simulations show that moderate FI pressure suppresses transport via zonal flow (ZF) shear, whereas strong FI drive weakens ZFs and enhances transport by destabilizing FI-BAEs. These results highlight the dual role of FIs in regulating turbulence and offer insight into multiscale transport physics relevant for high-performance plasmas.

## I. INTRODUCTION

Turbulent transport remains a central challenge in magnetically confined fusion plasmas, as it significantly degrades confinement performance. It is primarily driven by drift-wave microturbulence, which arises from temperature and density gradients and often exceeds neoclassical predictions by orders of magnitude [1,2]. Theoretically and numerically, heat and particle transport in typical tokamak plasmas is known to be dominated by ion temperature gradient (ITG) modes and trapped electron modes (TEM), while electron temperature gradient (ETG) modes play a critical role in electron heat transport [3-5]. While substantial progress has been made in understanding electrostatic microinstabilities in low-beta plasmas, electromagnetic (EM) turbulence is expected to dominate in high-beta regimes, such as those envisioned for ITER and DEMO.

Self-generated mesoscale zonal flows, including zero/low-frequency zonal flows and geodesic acoustic modes (GAMs), can regulate turbulence via nonlinear three-wave interactions and are particularly effective in the plasma core, often exceeding the shearing effects of mean flow gradients [6,7]. Another key player is the population of fast ions (FIs), generated via auxiliary heating such as neutral beam injection (NBI) or ion cyclotron resonance heating (ICRH). These energetic particles can stabilize ITG-driven turbulence through dilution effects, wave-particle resonance, and nonlinear interactions with zonal flows and micro-instabilities. Notably, FIs have been associated with the emergence of enhanced confinement regimes, such as the F-ATB mode observed in ASDEX Upgrade [8] and the FIRE mode in KSTAR [9]. Moreover, FIs may drive or enhance electromagnetic instabilities such as the kinetic ballooning mode (KBM), particularly in weak magnetic shear regions. Altogether, the complex interplay among turbulence, zonal flows, and fast ions plays a pivotal role in determining the transport characteristics and overall performance of fusion plasmas, and remains a critical focus of ongoing research.

Early investigations into the impact of FIs on turbulence were primarily driven by experiments and modeling on the JET tokamak. Linear gyrokinetic simulations have suggested that fast ions generated by ICRH may trigger internal transport barriers (ITBs) in high-performance hybrid scenarios. In such hybrid regimes [10], a clear correlation has been observed between improved confinement and elevated fast-ion content, which was later linked to a fast-ion-induced stabilization mechanism of EM ITG modes within gyrokinetic frameworks [11,12]. This mechanism was initially identified in numerical studies of dedicated JET ion heat transport experiments, which measured significant reductions in ion heat stiffness under on-axis ICRH and NBI heating [13,14,15]. (Here, stiffness is defined as the sensitivity of the gyroBohm-normalized heat flux to variations in the logarithmic ion temperature gradient.) In parallel, Citrin et al. numerically uncovered a fast-ion-enhanced nonlinear EM stabilization mechanism, capable of suppressing ITG turbulence beyond predictions from linear theory.

Extensive theoretical, numerical, and experimental studies have demonstrated that fast ions can exert a profound yet nontrivial influence on microturbulence and transport in magnetically confined plasmas. On the stabilizing side, fast ions modify equilibrium and fluctuation properties through pressure-gradient effects, dilution of thermal species, electromagnetic stabilization, and enhanced zonal-flow activity, leading to a reduction of ion-scale turbulence such as ITG and, in some regimes, TEM modes. Conversely, fast ions may also resonate with microinstabilities or Alfvénic fluctuations, driving fast-ion–induced modes and altering nonlinear saturation pathways, thereby weakening zonal flows and increasing turbulent transport. This coexistence of stabilizing and destabilizing mechanisms, and their strong dependence on fast-ion pressure gradients, phase-space distribution, and plasma β, has been identified as a key element in understanding turbulence regulation in high-performance and burning plasma regimes[16, 17].

On the other hand, the nonlinear coupling between fast-ion-driven modes and zonal flows represents a key issue in the study of burning plasmas. Several mechanisms have been identified, including forced drive during the linear growth phase of fast-ion modes [18,19], slower-time-scale modulational instabilities [20,21], and the excitation of geodesic acoustic modes (GAMs). Fast-ion modes are capable of generating zonal flows with both fine-scale and mesoscale structures [22], and are also expected to modify the residual zonal flow level through mechanisms associated with temperature anisotropy [23,24]. The consequences of such modifications for ITG saturation levels remain to be fully understood. Recently, global gyrokinetic simulations have been performed to investigate the interaction between ITG turbulence and beta-induced Alfvén eigenmodes (BAEs) [25]. These results indicate that, within the specific parameter regime explored, ITG modes are not stabilized, whereas kinetic ballooning modes (KBMs) remain robustly unstable. However, these simulations were not flux-driven. Building upon this context, this letter employs gyrokinetic simulations to investigate the nonlinear, multiscale impact of fast-ion pressure gradients on turbulent transport in high-β plasmas. Our work aims to uncover the hierarchical regulation mechanisms by which fast ions influence microturbulence and large-scale transport dynamics. These findings provide new insight into multiscale coupling in burning fusion plasmas and offer theoretical and computational support for achieving high-confinement regimes in tokamak operation and future reactor design.

## II. Simulation setup

The simulations presented in this work are based on a recent high-β discharge from the HL-2A tokamak. HL-2A is a medium-sized deuterium-fueled tokamak typically operated in limiter or lower single-null configurations, with a major radius of R=1.65m and a minor radius of r=0.4m. We focus on discharge #25803, which exhibits a well-developed ITB. This typical ITB experiments are performed in deuterium plasmas with plasma current Ip ~150 kA, toroidal magnetic field Bt ~ 1.37 T and central line-averaged density $\bar{n}_e$≈1.2 × $10^{19}$ m$^{-3}$. Figure 1 shows a representative ITB experimental result from this discharge. A 32-channel fast electron cyclotron emission (ECE) diagnostic system provides measurements of the electron temperature Te, with temporal and spatial resolutions of up to 0.8 μs and 1 cm, respectively [26]. The electron density ne profile is reconstructed from a formic acid (HCOOH) laser interferometer using Abel inversion techniques [27], and all profiles are mapped onto magnetic flux coordinates. The safety factor profile $q(\rho)$, shown in Fig. 1(a), is obtained using the kinetic EFIT reconstruction module within the OMFIT integrated modeling framework [28]. Due to off-axis neutral beam injection (NBI), the plasma core exhibits a weak magnetic shear configuration, which is conducive to the excitation of kinetic ballooning modes (KBMs) observed later in this study.The FI pressure and density profiles, shown in Figs. 1(c) and 1(d), are computed using the NUBEAM module embedded within the ONETWO transport solver in OMFIT [29].

*Contact author: lijc@szu.edu.cn

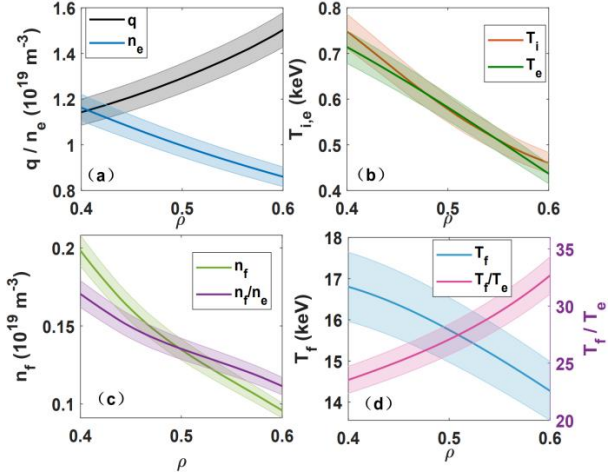

Figure 1. Profiles of #25803 at t = 662 ms. (a) **q** (black) and $n_e$ (blue), (b) temperatures of $T_i$ (red) and $T_e$ (green), (c) Fast ion density $n_f$ (green) and its ratio to $n_e$ (purple), (d) Fast temperature $T_f$ (blue) and its ratio to $T_e$ (pink).

The impact of fast ions on turbulence is investigated using the gyrokinetic code GENE. The input parameters for the simulations, derived from the equilibrium and profiles shown in Fig. 1, are summarized in **Table 1**. Prior to the main simulations, grid convergence tests were performed to ensure numerical accuracy. In the nonlinear GENE simulations, the poloidal box size is set to $L_y \sim 209\,\rho_s$, and the turbulent heat flux spectra peak in the range $0.01 < k_y\rho_s < 0.35$, indicating that the chosen domain is sufficiently large to capture the relevant modes. The radial box width is: $L_X \sim 136\,\rho_s$, discretized with 128 grid points, yielding a radial resolution of $\Delta r \sim 1.06\rho_s$. Despite the smaller radial width in the GENE simulations, the turbulent eddy structures observed in all simulations remain well below the domain size, confirming the adequacy of spatial resolution. The velocity space resolution was lowered to nv0 = 32 and nw0 = 16.

Table 1. Dimensionless parameters utilized in the simulations for the cases of with and without the contribution from fast ions.

|  | $\rho$ | $\hat{s}$ | $R/L_{ne}$ | $R/L_{ni}$ | $R/L_{Te}$ | $R/L_{Ti}$ | $T_i/T_e$ |
|---|---|---|---|---|---|---|---|
| W/O FI | 1.292 | 0.699 | 7.78 | 7.78 | 12.35 | 13.7 | 0.99 |
| W FI | 1.292 | 0.699 | 7.78 | 6.59 | 12.35 | 13.7 | 0.99 |
|  | $n_f/n_e$ | $R/L_{nf}$ | $R/L_{Tf}$ | $T_f/T_e$ |  |  |  |
|  | 0.14 | 16.5 | 4.28 | 27.1 |  |  |  |

## III. RESULTS

Figure 2 displays the linear growth rates (a) and real frequencies (b) for HL-2A discharge #25803, comparing electromagnetic (EM) and electrostatic (ES) simulations with and without FIs. The wavenumber spectrum can be broadly divided into two regions: a low-$k_y$ range ($k_y\rho_s \ll 0.6$), hereafter referred to as the low-k domain, and a higher-$k_y$ domain ($0.6 < k_y\rho_s \lesssim 1.5$) where TEM instabilities become dominant. In Fig. 2(a), the solid blue and red curves denote the most unstable modes in the absence of fast ions, showing a broad growth rate peak centered around $k_y\rho_s \sim 0.35$. Linear ES simulations with and without fast ions are conducted under otherwise identical conditions. Compared to EM cases, neglecting magnetic fluctuations in ES simulations leads to a modest increase in growth rates, while the low-$k_y$ region ($k_y\rho_s < 0.3$) remains largely unaffected. Notably, the presence of fast ions strongly stabilizes ITG modes, leading to a substantial reduction in both linear growth rates and the associated nonlinear heat flux, as reported in Refs. [11–15]. The observed ITG stabilization arises mainly from the dilution of the thermal ion species, which reduces the effective ITG drive [16, 30]. Under the present simulation conditions, the criterion required for resonant fast-ion effects, namely $\eta_f = (R/L_{Tf})/(R/L_{nf}) > 1$, is not satisfied. Consequently, the fast-ion distribution does not support significant wave–particle resonant interactions. This explains why the ITG growth rates in Fig. 3(a) (see below) show little sensitivity to the fast-ion logarithmic temperature gradient, which would indeed be expected if resonance effects were dominant. In contrast, the linear growth rates of TEM modes exhibit only a weak dependence on the presence of fast ions in both EM and ES simulations. Thus, under the HL-2A discharge conditions studied here, the influence of fast ions on TEM-driven turbulence appears to be negligible.

*Contact author: lijc@szu.edu.cn

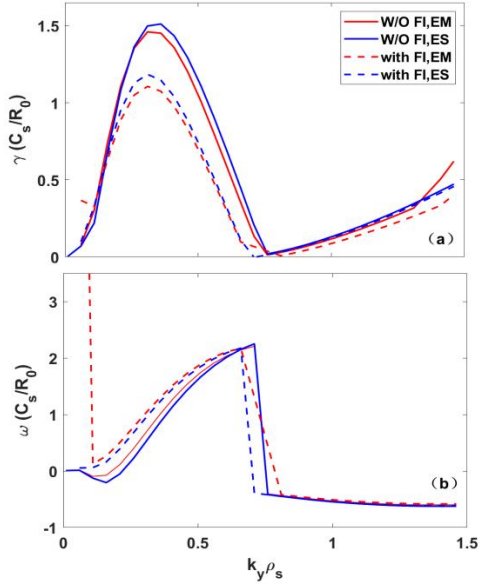

Figure 2. Linear growth rates (a) and frequencies (b) for the HL-2A 25803 discharge. A comparison among the electromagnetic and electrostatic cases both with fast ion and w/o fast ion.

Figures 3(a1-b2) present the linear growth rates and real frequencies from a scan over the fast-ion temperature gradient. The observed mode is driven by a combination of total thermal pressure and the gradient of the fast-ion pressure. It is evident that even a slight variation in the fast-ion pressure gradient induces a pronounced change in the growth rate in the low-$k_y\rho_s$ region. This transition-from lTG modes, primarily driven by thermal ion pressure gradients (as in the well-known AITG mechanism), to modes dominantly driven by fast-ion pressure gradients can be interpreted as a shift in the instability drive mechanism. Given that this latter mode resides within the BAE gap of the Alfvénic frequency spectrum, it is identified as a fast-ion-driven BAE (FI-BAE), consistent with the interpretation in Mazzi et al. [31]. Figures 3(b1-b2) show the linear growth rate and real frequency as a function of the electron $\beta_e$. At low $\beta_e$, ITG and TEM modes dominate. However, as $\beta_e$ increases, high-growth-rate modes emerge, corresponding to KBMs. Notably, KBMs are absent in the electrostatic limit ($\beta_e=0$). When $\beta_e=0.25\%$, low-$k_y\rho_s$ KBMs become unstable, while higher-$k_y\rho_s$ modes remain consistent with lTG behavior, as inferred from their real frequency signatures. In GENE, positive real frequency corresponds to propagation in the ion diamagnetic direction. As $\beta_e$ further increases, KBMs extend to higher $k_y\rho_s$, as shown in Fig. 3(a2). At the experimental value $\beta_e^{exp} = 0.12\%$, the dominant microinstabilities are long-wavelength lTG and TEM modes, indicating that the plasma resides in an lTG-TEM hybrid turbulence regime under these conditions.

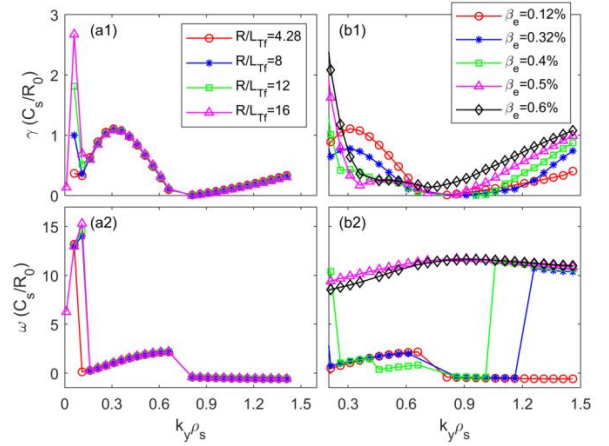

Figure 3. the growth rate (a1-b1) and mode frequency (a2-b2) for the scan over the temperature gradient of fast ion and the βe of the HL-2A #25803 discharge.

Figure 4(a-c) illustrates the mode structures corresponding to the experimental value $\beta_e^{exp}=0.12\%$, $\beta_e=0.4\%$, and $\beta_e=0.6\%$, respectively. Consistent with theoretical predictions [32], both the ES potential ϕ and the parallel magnetic vector potential $A_\parallel$ for KBM and ITG modes exhibit ballooning and tearing-parity structures along the field line. Due to differences in wavenumber, the lTG mode appears more spatially localized near the outboard midplane (θ= 0). Furthermore, the presence of oscillatory side bands-characteristic of such mode structures may weaken the nonlinear coupling between ZFs and turbulence. A similar feature is also observed in TEM-driven turbulence, where the linear eigenmode tends to extend more broadly along the ballooning angle, as reported in Ref. [33]. Furthermore, nonlinear simulations for HL-2A discharge #25803 were performed using the GENE code. Figure 4(d1-d2) shows the time evolution of the ion electrostatic heat diffusivity and electrostatic (ES) flux for three different values of the fast-ion pressure gradient. As the fast-ion pressure gradient increases, both the electron heat diffusivity and ES flux rise significantly, indicating an enhancement in turbulent transport. This increased transport is attributed to the destabilization of a fast-ion–driven

*Contact author: lijc@szu.edu.cn

BAE mode, induced by elevated total plasma pressure.

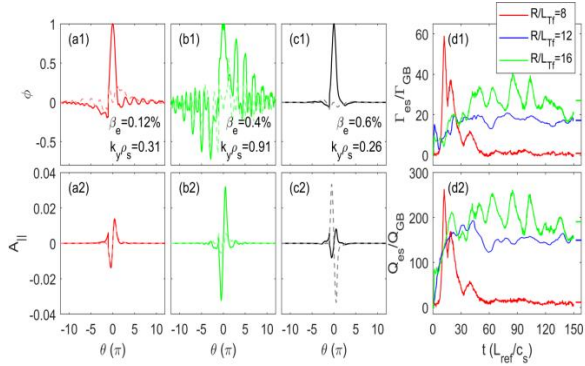

Figure 4. (a-c)Typical mode structure of the TEM, KBM, and ITG, respectively. Time evolutions of (d1) electrostatic heat diffusivity and (d2) electrostatic heat flux.

Figure 5 displays the ZF structures of the electrostatic potential $\tilde{\phi}$ and density perturbations for the three fast-ion gradient cases. When the fast-ion temperature gradient $R/L_{Tf}$ is large, the radial correlation length in the ES potential is relatively short, and no clear coherence is observed along the field line. In contrast, for small $R/L_{Tf}$, pronounced long-range correlations emerge in the binormal direction, accompanied by rippling structures in the radial direction-indicative of strong shear, as seen in Fig. 5(a1). In this regime, large-scale turbulent eddies are broken up and scattered to higher wavenumber domains, thereby suppressing ion heat transport and improving confinement.

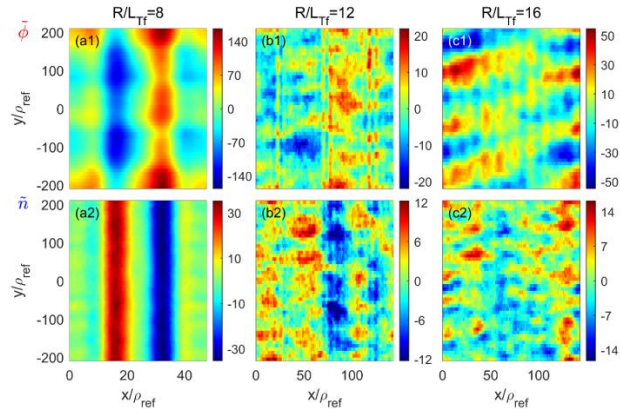

Figure 5. Spatial structure of electrostatic potential for three vales of the temperature gradient of fast ion.

It should be noted that in this study we focused on scenarios where KBM and TAE modes are weakly stable or unstable, for which the flux-tube model is adequate. However, we acknowledge that for stronger modes or more significant energetic particle effects on turbulence, the flux-tube model has limitations, as it may not fully capture global dynamics. As highlighted by Di Siena et al. [34], radially global simulations are essential for accurately modeling stronger KBM and TAE modes. While our current analysis does not employ global simulations, we recognize their importance and plan to explore this approach in future work to provide a more comprehensive understanding of energetic particle interactions.

## IV. CONCLUSIONS

In summary, we have employed gyrokinetic simulations using the GENE code to investigate the impact of FIs pressure gradients on turbulence in high-$\beta$ plasmas based on HL-2A discharge #25803. Linear simulations reveal that fast ions significantly reduce the growth rate of lTG modes,particularly in the low-$k_y\rho_s$ regime, while having negligible influence on TEM instabilities. This stabilizing effect is attributed to the thermal-ion dilution, which occurs when the FI magnetic drift frequency overlaps favorably with the ITG mode frequency in phase space. At higher fast-ion temperature gradients, a transition from lTG-dominated turbulence to a FI-driven BAE regime is observed, supported by mode frequency alignment with the BAE gap and structure characteristics consistent with ballooning parity. $\beta_e$ scans further confirm the emergence of KBM instabilities at higher 8., whereas lTG-TEM hybrid turbulence dominates at the experimental value $\beta_e^{exp} \approx 0.12\%$. Nonlinear simulations show that increasing FI pressure gradient can destabilize the FI-BAE mode, leading to enhanced electrostatic flux and heat diffusivity, consistent with pressure-driven transport enhancement. Analysis of the electrostatic potential structures reveals that for weaker FI drive, ZFs are strong and capable of shearing turbulent eddies, thereby suppressing transport. In contrast, for stronger FI drive, ZFs are weakened and radial correlation lengths increase, allowing large-scale eddies to form and enhance transport. These results highlight the dual role of fast ions in turbulence regulation: they can both stabilize microinstabilities through wave–particle interactions and destabilize macroscopic electromagnetic modes that degrade confinement. Our findings provide new insight into multiscale

*Contact author: lijc@szu.edu.cn

fast-ion–turbulence interactions, with implications for achieving optimized transport regimes in high temperature plasmas.

## ACKNOWLEDGMENTS

This work was partly supported by the National MCF Energy R&D Program (Grant Nos. 2024YFE03190004 and 2024YFE03190001), National Natural Science Foundation of China (Grant Nos. 12475215, 12405257 and 12405254), Shenzhen Municipal Collaborative Innovation Technology Program-International Science and Technology (S&T) Cooperation Project (GJHZ20220913142609017), and the Shenzhen Science and Technology Program (ZDSYS20230626091501002).

*Contact author: lijc@szu.edu.cn

*Contact author: lijc@szu.edu.cn